\documentclass[aps,physrev,preprint,groupedaddress]{revtex4-2}

\usepackage[english]{babel}
\usepackage{graphicx} % inclusion des figures
\usepackage{caption}
\usepackage{subcaption}
\usepackage[separate-uncertainty=true]{siunitx}
\usepackage{physics}
\usepackage[dvipsnames]{xcolor}
\newcommand{\refun}{\textcolor{black}}
\newcommand{\refdeux}{\textcolor{black}}
\newcommand{\reftrois}{\textcolor{black}}
\newcommand\Rey{\mbox{\textit{Re}}}

\begin{document}

%%%% Article title to be placed here
\title{Formation and fragmentation of nylon fibre aggregates}

\author{L. Gey}
\author{P. Le Gal}
\author{G. Verhille}
\email[G. Verhille mail :]{gautier.verhille@cnrs.fr}
%\homepage[]{Your web page}
%\thanks{}
%\altaffiliation{}
\affiliation{Aix Marseille Univ., CNRS, Centrale Méditerranée, IRPHE, F-13013 Marseille, France}
%\author{%%%% Author details
%L. Gey$^{1}$, P. Le Gal$^{1}$ and G. Verhille$^{1}$}
%
%%%%%%%%%% Insert author address here
%\address{$^{1}$Aix Marseille Univ., CNRS, Centrale Méditerranée, IRPHE, F-13013 Marseille, France}
%
%%%%% Subject entries to be placed here %%%%
%\subject{Physics, Fluid mechanics}
%
%%%%% Keyword entries to be placed here %%%%
%\keywords{Aggregation, Fragmentation, Fibre network}
%
%%%%% Insert corresponding author and its email address}
%\corres{G. Verhille\\
%\email{gautier.verhille@cnrs.fr}}
\date{\today}
%%%% Abstract text to be placed here %%%%%%%%%%%%
\begin{abstract}
The aggregation process of fibers suspended in a fluid remains an open question, despite its critical role in numerous natural and industrial processes.
This paper presents an experimental setup designed to create fiber aggregates in a turbulent flow.
The phase diagram for the aggregation states is established as a function of the experimental parameters, and the statistical properties of the resulting aggregates are characterized.
Additionally, the fragmentation process is investigated, and a model is proposed to describe the fragmentation time of the aggregates, which is subsequently validated through comparison with experimental results.
\end{abstract}
%%%%%%%%%%%%%%%%%%%%%%%%%%%

%\rsbreak
\maketitle
%%%%%%%%%% Insert the texts which can accomdate on firstpage in the tag "fmtext" %%%%%

\section{\label{SecIntro} Introduction}
%%%% Insert A head here

\reftrois{The aggregation of fibers in turbulent flows plays a critical role in a wide range of natural and industrial processes. 
In nature, an example is the formation of aegagropilae, which are fiber aggregates resulting from the decomposition of \textit{Posidonia oceanica}~\cite{Russell_1893,cannon_experimental_1979,sanchez-vidal_seagrasses_2021,verhille_structure_2017}. 
Although these structures have been observed for many years, the precise mechanisms underlying their formation remain poorly understood. 
Aegagropilae are of interest not only from a fundamental scientific perspective~\cite{cannon_experimental_1979,verhille_structure_2017}, but also due to their ecological significance within the Mediterranean ecosystem~\cite{telesca_seagrass_2015}.}

\reftrois{However, fiber aggregation is not unique to this natural phenomenon. 
Gaining a deeper understanding of this process could yield valuable insights across various industries. 
For instance, fiber aggregation is central to the papermaking process~\cite{lundell_fluid_2011} as paper is essentially a fiber network formed from a suspension of wood fibers. 
Similarly, challenges in environmental management—such as plastic and textile pollution or wastewater treatment—also involve fiber aggregation mechanisms. 
Domestic wastewater from washing machines can release more than 1900 fibers per wash. 
Many of these fibers—whether polyester, acrylic, or cotton—can ultimately reach marine environments~\cite{browne_accumulation_2011}. 
In addition, abandoned fishing gear contributes an estimated \SI{1e5}{\tonne} of waste annually, posing serious threats to marine ecosystems~\cite{kuczenski_plastic_2022}.}

\reftrois{More broadly, fiber aggregation in turbulent flows lies at the intersection of several fundamental questions in physics.
Understanding and modeling this process begins with the study of general aggregation dynamics, which are often investigated using spherical particles~\cite{shaw_particle-turbulence_2003,wilkinson_caustic_2006,zaichik_two_2003,Bec2024,pumir_collisional_2016}. 
However, fibers differ significantly: as elongated, and sometimes deformable objects, they interact with turbulence in unique ways that must be accounted for~\cite{brouzet_flexible_2014,arguedas-leiva_elongation_2022}. 
Furthermore, studying the cohesion of fiber aggregates requires examining fiber networks, linking the microscopic properties of individual fibers to the macroscopic mechanical behavior of the entire structure~\cite{picu_mechanics_2011,van_wyk_20note_1946,gey_structural_2025}.}

This work is built on previous studies of fibre aggregation in flows. 
Soszynski et al.~\cite{soszynski_elastic_1988, soszynski_elastic_19882} observed elastic interlocking of nylon fibres using a rotating tilted cylinder filled with a suspension of fibres in water. 
Kerekes et al.~\cite{kerekes_characterization_1992} introduced the crowding number as a key parameter for characterizing aggregation:
\begin{equation} 
N_\text{C} = \frac{2}{3}N\frac{V_\text{f}}{V_\text{tot}}\qty(\frac{L}{d})^2 
\label{crowding}
\end{equation}
Here, $N$ is the number of fibres in the volume $V_\text{tot}$, $V_\text{f}$ is the volume of a single fibre of length $L$ and diameter $d$.

\begin{figure}[htbp]
    \includegraphics[width=1\linewidth]{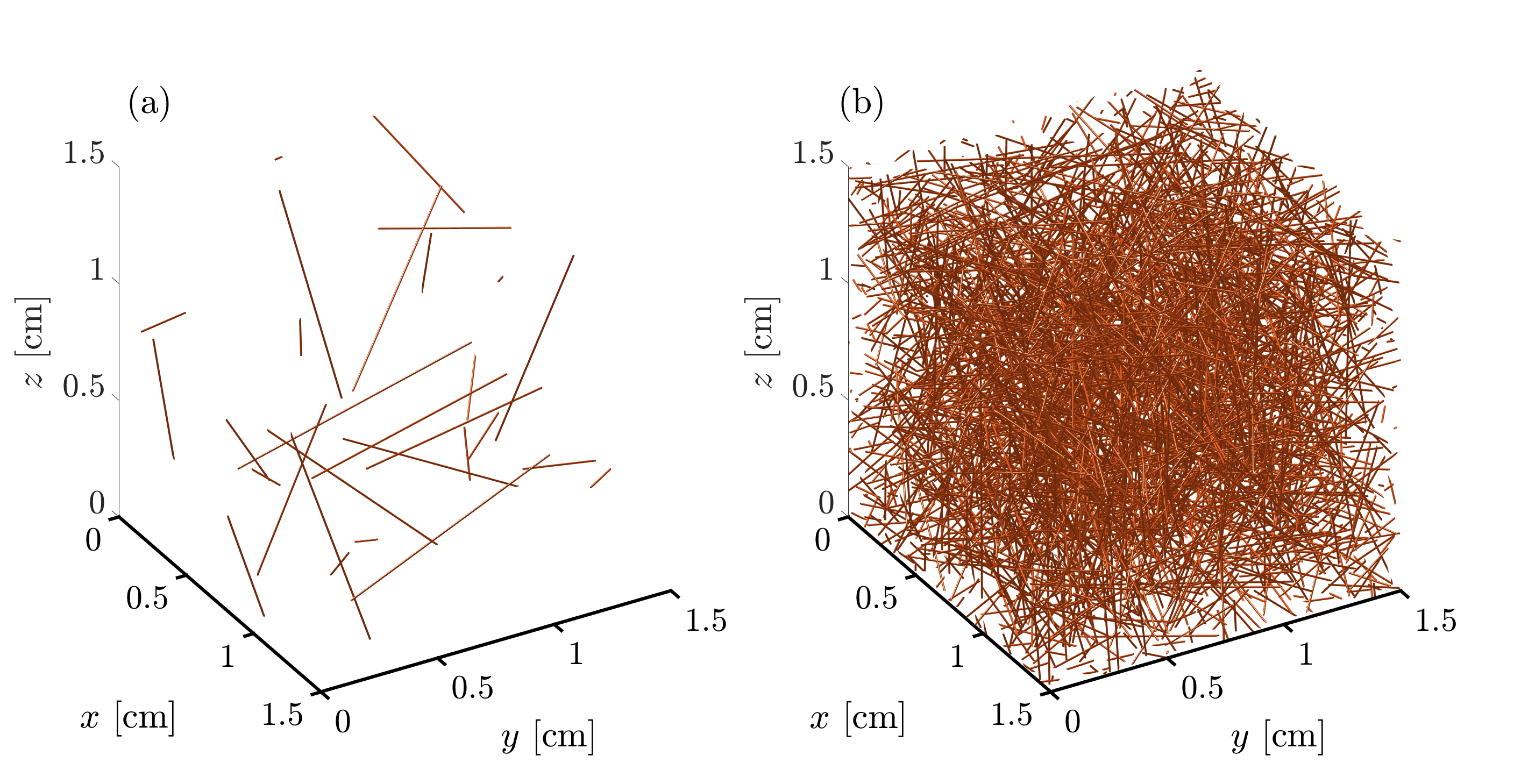}
    \caption{Example of randomly distributed straight fibres for different crowding number.
    The aspect ratio of the fibres is $\Lambda = L/d = 133$. (a) $N_\text{C} = 1$ and (b) $N_\text{C}=L/d$. To simulate the fibres, a numerical code from~\cite{Ayad_Al_Rumaithi} is used.}
    \label{fig:crowding}
\end{figure}

The crowding number quantifies the number of fibres within a sphere of diameter equal to the fibre length and is said to be critical for determining whether the fibre collision rate is high enough to enable aggregation.
Kerekes et al. \cite{kerekes_characterization_1992} identified 3 regimes.
(i) If $N_\text{C}<1$, the fibres does not interact, and the system is classified as dilute.
(ii) If $1<N_\text{C}<L/d$, the system is semi-dilute.
Aggregates may form if $L/d>50$, but they remain loose.
(iii) If $N_\text{C}>L/d$ and $L/d>50$, the fibres frequently interact, leading to aggregation in the concentrated regime.
To have a better insight of the concentration associated with these regimes, examples of randomly distributed fibres are presented on Figure~\ref{fig:crowding}.

Numerical simulations have further demonstrated that aggregation can occur even in the absence of attractive forces between fibres \cite{schmid_simulations_2000}. 
These studies have also quantified key parameters such as friction between fibres and curvature \cite{switzer_flocculation_2004}.

For the present study, the term aggregate refers to cohesive structures that can be removed from a flow while maintaining their shape.
\refdeux{The novelty of this article lies in the use of a turbulent setup that enables the formation of dense and cohesive fibre aggregates. The characterization of the domain of existence of these aggregates shows the crutial role of the turbulence intensity in the aggregation process. This observation extends the work of Soszynski et al.~\cite{soszynski_elastic_1988}.
Moreover, a new model for the fragmentation of the aggregates is proposed, based on the reorganization of the network during the fragmentation process.}
\refdeux{In Section~\ref{SecExpSetup}, an experimental setup based on a von Kármán turbulent flow is presented, which enables the formation of dense and cohesive aggregates.
Section~\ref{Formation} describes the phase diagram of aggregation as a function of the volumic fraction of the suspension $\varphi$ and the turbulence intensity.
Aggregation is only observed within a narrow range of parameters. 
The formed aggregates are found to undergo compaction during their formation process.
Finally, in Section~\ref{secFrag}, the fragmentation process is investigated with a specific insight into the fragmentation time of the aggregates. 
A model is proposed based on the reorganization of the fibers during their collisions with the discs.
This model shows a good agreement with the experimental data presented in this article.
}

%%%%%%%%%%%%%%%%%%%%%%%%%%%%%%
%%%%%%%%%%%%%%%%%%%%%%%%%%%%%%
%%%%%%%%%%%%%%%%%%%%%%%%%%%%%%
\section{\label{SecExpSetup}Experimental setup}

The experimental setup used for the formation of nylon fibre aggregates has been previously described in~\cite{gey_structural_2025}. 
Here, the main characteristics of the setup are summarized.

\begin{figure}[htbp]
    \includegraphics[width=1\linewidth]{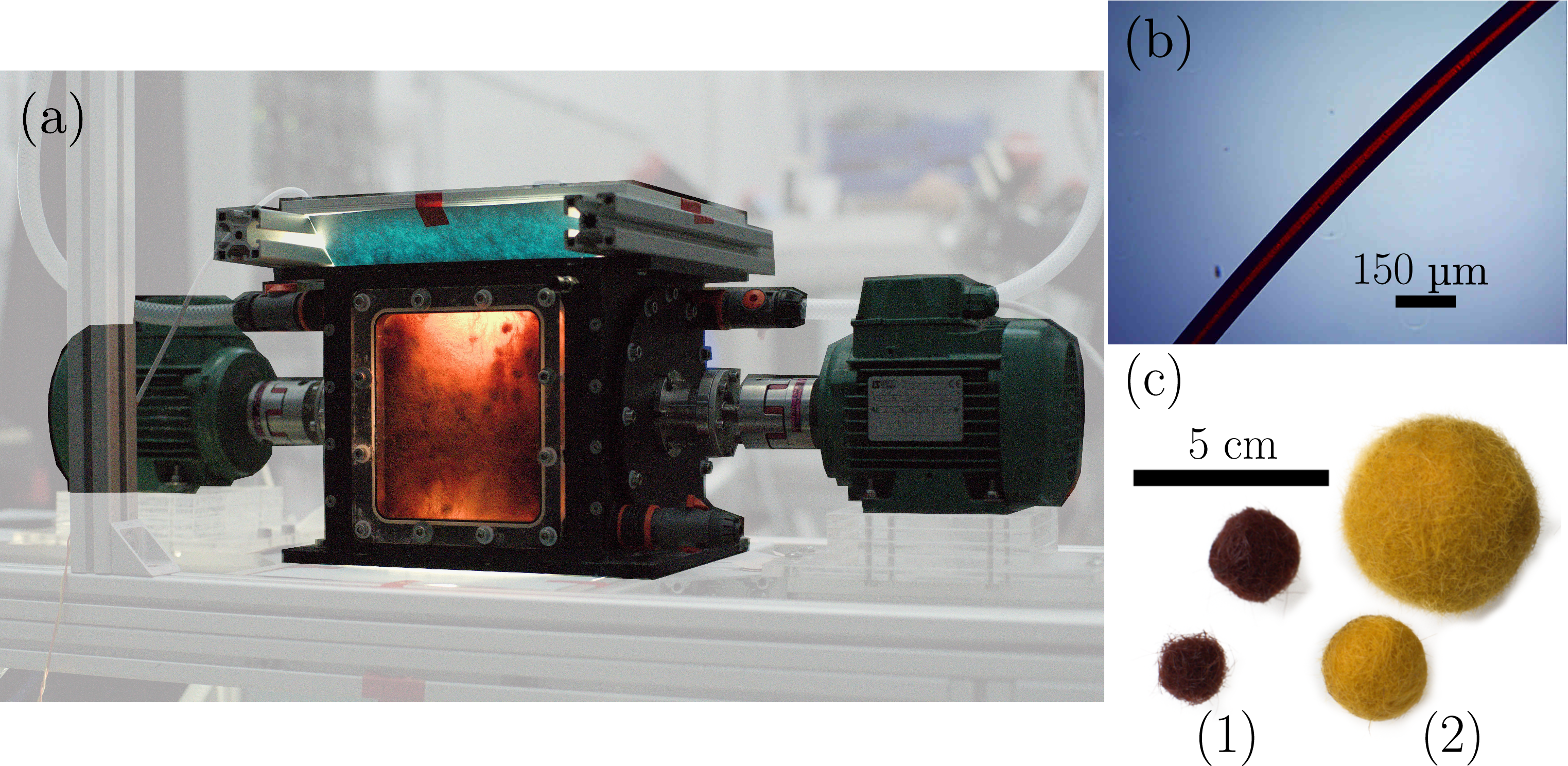}
   \caption{(a) Image of the experimental setup. (b) Microscope image of a fibre from set 1. The magnification is given by the black line whose length is \SI{150}{\micro\meter} (c) Example of aggregates generated with the experimental set-up. (1) and (2) correspond to aggregates obtained with nylon fibres from set 1 and 2 respectively.}
    \label{fig:Setup}
\end{figure}

The system consists of a von Kármán turbulent flow within a \SI{20}{\centi\meter} cubic tank filled with water, as shown in Figure~\ref{fig:Setup}.
\refun{The von Kármán turbulent flow has been widely used in various applications to generate flows at high Reynolds numbers~\cite{bourgoin_experimental_2006,saint-michel_probing_2014,volk_dynamics_2011}.}
The flow is generated by the rotation of two smooth disks \refun{of radius $R_\text{d}=$}$\SI{8}{\centi\meter}$ rotating in opposite directions.
\refun{Defining the integral Reynolds numbers as $\Rey = 2\pi f R_\text{d}^2/\nu$, where $f$ is the motor frequency and $\nu$ is the kinematic viscosity, the flow is considered fully turbulent if $\Rey > \SI{3.3e3}{}$~\cite{ravelet_supercritical_2008}. 
However, the turbulence is known to be inhomogeneous and anisotropic, with counter-rotating azimuthal cells~\cite{machicoane_particules_2014}. 
In the turbulent regime, the classical scaling laws from Kolmogorov’s theory of turbulence are recovered at the center of the tank~\cite{volk_dynamics_2011}.}

In the absence of fibres, the flow characteristics were analysed using Particle Image Velocimetry (PIV).
For a motor frequency $f$ ranging from \qtyrange{2.5}{21.5}{\Hz}, \refun{the integral Reynolds number ranges from \qtyrange{1e5}{8.6e5}{} and is then fully turbulent.}
The energy dissipation rate $\epsilon$ at the central region of the tank follows a $f^3$ scaling law, with values ranging from \qtyrange{1e-4}{2e-1}{\square\meter\per\cubic\second} \refun{and the Kolmogorov length scale $\eta$ ranges from \qtyrange{4.7e-2}{3.2e-1}{mm}.}

To form fibre aggregates, commercial nylon fibres from DCAFlock are used. 
These fibres are often used for special effects to simulate textures.
The characteristics of the flow is assumed to stay the same in the presence of fibres. 
Indeed, numerically, it has been found~\cite{olivieri_fully_2022} that the main characteristics of the flows such as the micro-scale Reynolds number or the turbulent kinetic energy are not much impacted by the presence of fibres up to a mass fraction of \num{2e-1}. Here the mass fraction does not exceed \num{2.5e-2}.
Moreover, no difference was observed in the evolution of the temperature (related to $\epsilon$) with or without fibres (not shown).
These fibres are slightly denser than water, with a density of $\rho = \SI{1140 \pm 10}{\kilogram\per\cubic\meter}$, slightly bent with a mean curvature $\kappa$, and have a Young's modulus estimated at $E \sim \SI{3e9}{\pascal}$. 
The characteristics of the different fibre sets are presented in Table~\ref{TableDim}.

\begin{table}[!h]
\centering
\caption{\label{TableDim} Fibres dimensions.}
\begin{tabular}{cccc}
\hline
 & Set 1& Set 2 & Set 3\\
        Fibre length $L$ [\SI{}{\centi\meter}] &  \num{1 \pm 0.02}& \num{0.8 \pm 0.02}&\num{0.6 \pm 0.02}\\
        Fibre diameter $d$ [\SI{}{\micro\meter}]& \num{75 \pm 2}& \num{75 \pm 2}& \num{80 \pm 2}\\
        Mean normalized curvature $\kappa L$ & \num{1.15\pm0.05}& \num{0.81\pm0.03}& \num{0.62\pm0.02} \\\hline
\end{tabular}
%\vspace*{-4pt}
\end{table}

The experimental procedure begins with a suspension of fibres, which is stirred at $f=\SI{25}{\hertz}$ for \SI{10}{\minute} to homogenize the flow. 
Subsequently, the frequency is decreased to the desired test value, and the experiment is run for durations ranging from \qtyrange{1}{48}{h}. 
Examples of aggregates formed during the experiments are shown on Figure~\ref{fig:Setup}~(c). 
The structural and mechanical properties of the generated aggregates are presented in details in a separate study~\cite{gey_structural_2025}.
To observe aggregation, high volume fractions of fibres are required to reach a crowding number high enough~\cite{kerekes_characterization_1992}.
In the results shown here, the volume fraction for which aggregation has been observed is ranging from $\varphi =NV_\text{f}/V_\text{tot} = \qtyrange{7e-3}{2e-2}{}$.
As the fibres are opaque, direct visualization of the aggregation process during the experiment is not possible. 
Consequently, it is not feasible to observe proto-aggregates or the nucleation process. 
The aggregation process is then characterized through the structural properties of the aggregates.

During an experiment, the number of aggregates ranges from zero to approximately 500.
After an experiment, the aggregates are collected and dried.
The size and shape of the aggregates are measured using photographs.
The mean aspect ratio of the aggregates is approximately \num{0.93 \pm 0.01}, therefore they can be considered as spherical at first order.
The number of fibres in an aggregate $N$ is given by the ratio of its mass $m$ to the mass of a fibre $m_\text{f}$ and then known within an uncertainty of $5\%$.
%%%%%%%%%%%%%%%%%%%%%%%%%%%%%%%%%%%%%%%%%%%%%%%%%%%%%%%%%%%%
%%%%%%%%%%%%%%%%%%%%%%%%%%%%%%%%%%%%%%%%%%%%%%%%%%%%%%%%%%%%
%%%%%%%%%%%%%%%%%%%%%%%%%%%%%%%%%%%%%%%%%%%%%%%%%%%%%%%%
\section{\label{Formation} Formation of aggregates}
%%%%%%%%%%%%%%%%%%%%%%%%%%%%%%%%%%%%%%%%%%%%%%%%%%%%%%%%%
\subsection{\label{Existance} Phase diagram}

\begin{figure}[htbp]
    \includegraphics[width=1\linewidth]{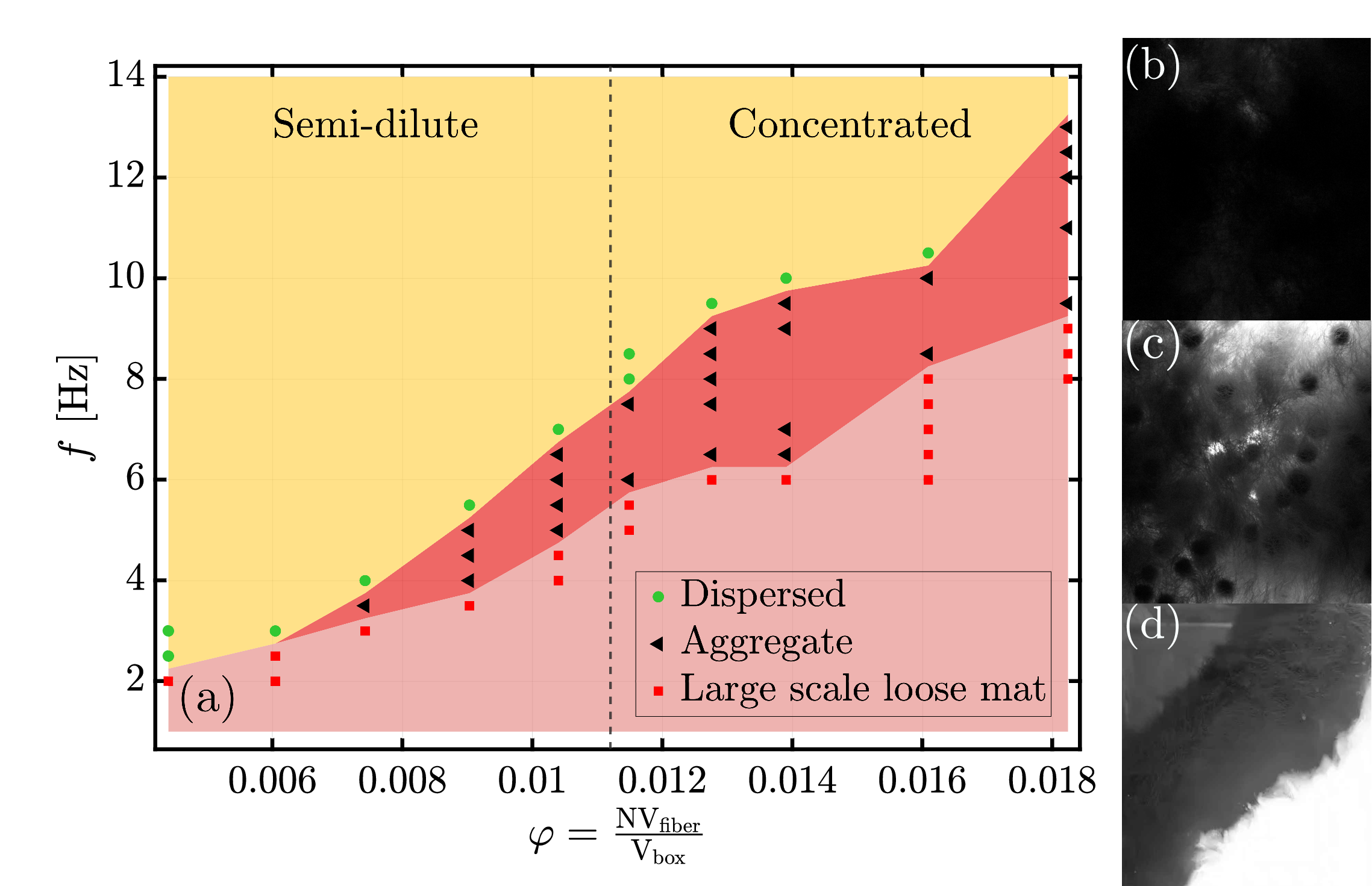}
    \caption{(a) Diagram of existence of the aggregation process for \SI{1}{\centi\meter} long fibres from set 1. The green circles correspond to the dispersed case: no highly cohesive structure are observed. What is seen by the camera when the tank is illuminated using a back-lighting and the flow is homogeneous as can be observed in snapshot (b). The black triangles correspond to the range of parameters where aggregation occurs as illustrated on (c). The red squares correspond to the cases where some or a majority of the fibres are whether settling or creating a loose pack at the scale of the tank and stopped moving as in (d). The dashed line correspond to $\varphi = \num{1.12e-5}$, the limit between the semi-dilute case and the concentrated case as defined by Kerekes et al.~\cite{kerekes_characterization_1992}.}
    \label{fig:existance}
\end{figure}

In the experiment there are several control parameters: the rotational frequency of the disc $f$, the volume fraction of fibres $\varphi$, the mechanical properties of the fibres (density, friction coefficient, sizes), and the properties of the fluid (density, viscosity). 
Here, the focus is placed on the first two of them in order to construct the phase diagram of the system (see Figure~\ref{fig:existance}) using fibres from set 1.

The phase diagram shows that if the flow intensity is high enough or the volume fraction is low enough, the hydrodynamic forces keep the fibres dispersed.
Conversely, if the frequency is low or the volume fraction is high, the fibres either settle or form a non-moving, loose packet at the scale of the tank.
The large-scale structures persisted even when salt were used to match the density of the fibers. 
Between these two extreme regimes, aggregates form.

Following Kerekes et al.~\cite{kerekes_characterization_1992}, one can compute the crowding number to compare it with the volume fraction at which the aggregates are observed.
In this approach the influence of the flow intensity is not taken into account.
The crowding number can be rewritten from Equation~\ref{crowding} as
$N_\text{C}=(2/3)\varphi\qty(L/d)^2$, where $\varphi$ is the volume fraction of the suspension. 
In our case, a dilute system is define by $\varphi < {3}/{2} \qty({d}/{L})^2 = \SI{8.4e-5}{}$, the concentrated regime for $\varphi> {3d}/{2L} = \SI{1.12e-2}{}$ and between these limits lies the semi-dilute regime.
This criterion provides a reasonable order of magnitude for the onset of aggregation, as shown in Figure~\ref{fig:existance}. 
However, by definition it cannot account for the influence of turbulence. % the frequency.
Focusing on the upper boundary between the formation of aggregates and the dispersed state, it is observed that higher concentrations allow aggregates to form at higher frequencies. 
This phenomenon can be explained as follows:
at any given moment, fluctuations in the flow may lead to collision of fibres and the formation of loose cohesive structures. 
Turbulence tends to fragment these structures.
However, if the fibre concentration is sufficiently high, additional fibres can aggregate and reinforce the structure before it is fragmented by turbulence, enabling aggregate growth.
Therefore, the data show that the prediction by Kerekes et al~\cite{kerekes_characterization_1992} should be adjusted based on the turbulence rate of the flow that agitates the fibres.

The same trend is observed for \SI{8}{\milli\meter}-long fibres from set 2.
However, we were unable to form aggregates with the \SI{6}{\milli\meter}-long fibres from set 3. 
Two factors may explain this observation.
First the fibre morphology: 
At rest, the fibres are not straight and their curvature facilitate the aggregation process~\cite{switzer_flocculation_2004}.
As shown on Table~\ref{TableDim}, $\kappa L$ is decreasing from set 1 to 3 reducing as a consequence their possible steric interactions. %steric interactions.  
Fibre flexibility may also play a crucial role in maintaining the structural integrity of aggregates \cite{soszynski_elastic_1988}. 
As flexibility decreases from set 1 to set 3, the ability to sustain cohesive aggregates diminishes.
Indeed, the bending of the fibres can be responsible of elastic interlocking~\cite{soszynski_elastic_1988}: the force due to the bending of the fibres combined with friction between the fibres can help the cohesion of the aggregate.
Following the scaling proposed by Brouzet et al. \cite{brouzet_laboratory_2021}, the curvature of a free fibre in a turbulent flow can be expressed as:
\begin{equation}
    \kappa \sim \frac{\qty(\eta \rho \epsilon)^{1/2}L^3}{EI}
    \label{eq:kappa}
\end{equation}

where $\eta$ is the dynamic viscosity of the fluid, $\rho$ is the fluid density, $\epsilon$ is the dissipation rate per unit mass, $E$ is Young’s modulus, and $I$ is the moment of area of the fibre cross-section.
According to this scaling, the predicted curvature is 6.3 times higher for the fibres in set 1 compared to those in set 3. 
This reduced curvature for fibres from set 3 likely impedes the interlocking process required for aggregation as the cohesion of fibre network can indeed be due to elastic interlocking of fibres~\cite{soszynski_elastic_1988}.
However, note that this elastic energy storage in the aggregate have not been detected on the measurements of curvature of the fibres done on the aggregates using X-ray tomography~\cite{gey_structural_2025}.

For the following of this study, all data and analysis will focus on experiments conducted with fibres from set 1.

The distribution of aggregate sizes and the number of fibers within the aggregates are characterized as functions of the input parameters of the experimental setup.

%%%%%%%%%%%%%%%%%%%%%%%%%%%%%%%%%%%%%%%%%%%%%%%%%%%%%%%%%%%%
\subsection{\label{Characterisation} Characterisation of the aggregates}

\begin{figure}[htbp]
    \includegraphics[trim={3.2cm 0 4.5cm 0},clip,width=1\linewidth]{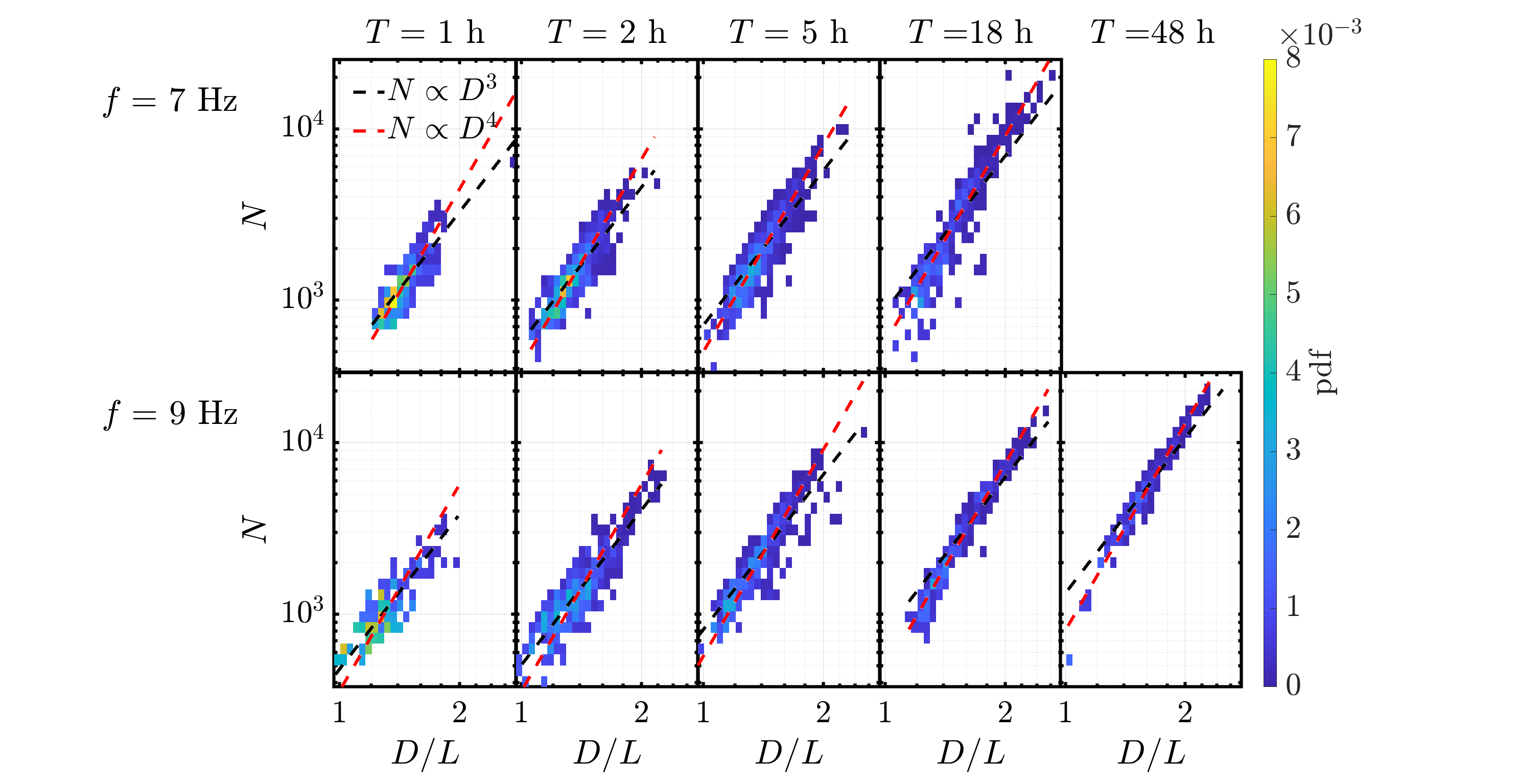}
    \caption{\refun{Joint} pdf of the normalized diameter $D/L$ and number of fibres $N$ of the aggregates. Each graph corresponds to a different experiment. In all the graphs, the initial volume fraction is $\varphi = \SI{1.32e-2}{}$. The first row corresponds to a forcing frequency of $ = \SI{7}{\hertz}$ and the second row to $f=\SI{9}{\hertz}$. The duration of the experiment increases from left to right.
    The black and red dashed lines represent the best fits of the experimental data assuming the scaling law $D^3$ (in black) and $D^4$ (in red). The best fits correspond to $N = \alpha (D/L)^4$ with $\alpha = \num{4.8(0.6)e2}$.}
    \label{fig:NRcol}
\end{figure}

In each experiment, the aggregates are collected and dried, then their sizes are measured and they are weighted.
This allow the determination, for each aggregate, of its diameter $D$ and its number of fibres $N$ (knowing the mass of a fibre).

The \refun{joint} distributions of $D/L$ and $N$ are shown on Figure~\ref{fig:NRcol}.
On this figure, each graph represents the probability distribution function (pdf) corresponding to a different experiment.
All experiments presented in this figure have the same initial volume fraction, $\varphi = \SI{1.32e-2}{}$, with fibres from set 1.
Only the duration $T$ of the experiment and the rotation frequency $f$ of the discs are varied.
If the aggregates would grow with a constant density, the number of fibre in an aggregate would be proportional to its volume, implying the scaling: $N\propto (D/L)^3$.
However, in the experimental data, the observed scaling is $N\propto (D/L)^4$, indicating that larger aggregates are denser than smaller ones.
This suggests that during the formation process, the aggregates undergo compaction.
This trend becomes more pronounced as the experiment runs for longer time.
However, since the aggregate cannot be compacted indefinitely, it is expected to eventually recover a scaling of at most $N\propto (D/L)^3$ over long timescales, as it is observed for aegagropilae~\cite{verhille_structure_2017} for example.
For percolated system, the exponent 3 is an upper limit at steady state~\cite{stauffer_scaling_1979}.
\reftrois{Note that, for clarity in the presentation of the data in Figure~\ref{fig:NRcol}, the fiber length $L$ is used as a typical length scale to normalize the aggregate diameter. 
However, this natural choice is not the only possible one. 
Other lenght, such as the fiber diameter $d$, or $l = (Ld^2)^{1/3}$, which accounts for the fiber volume, could also have been considered.}

\begin{figure}[htbp]
    \includegraphics[width=1\linewidth]{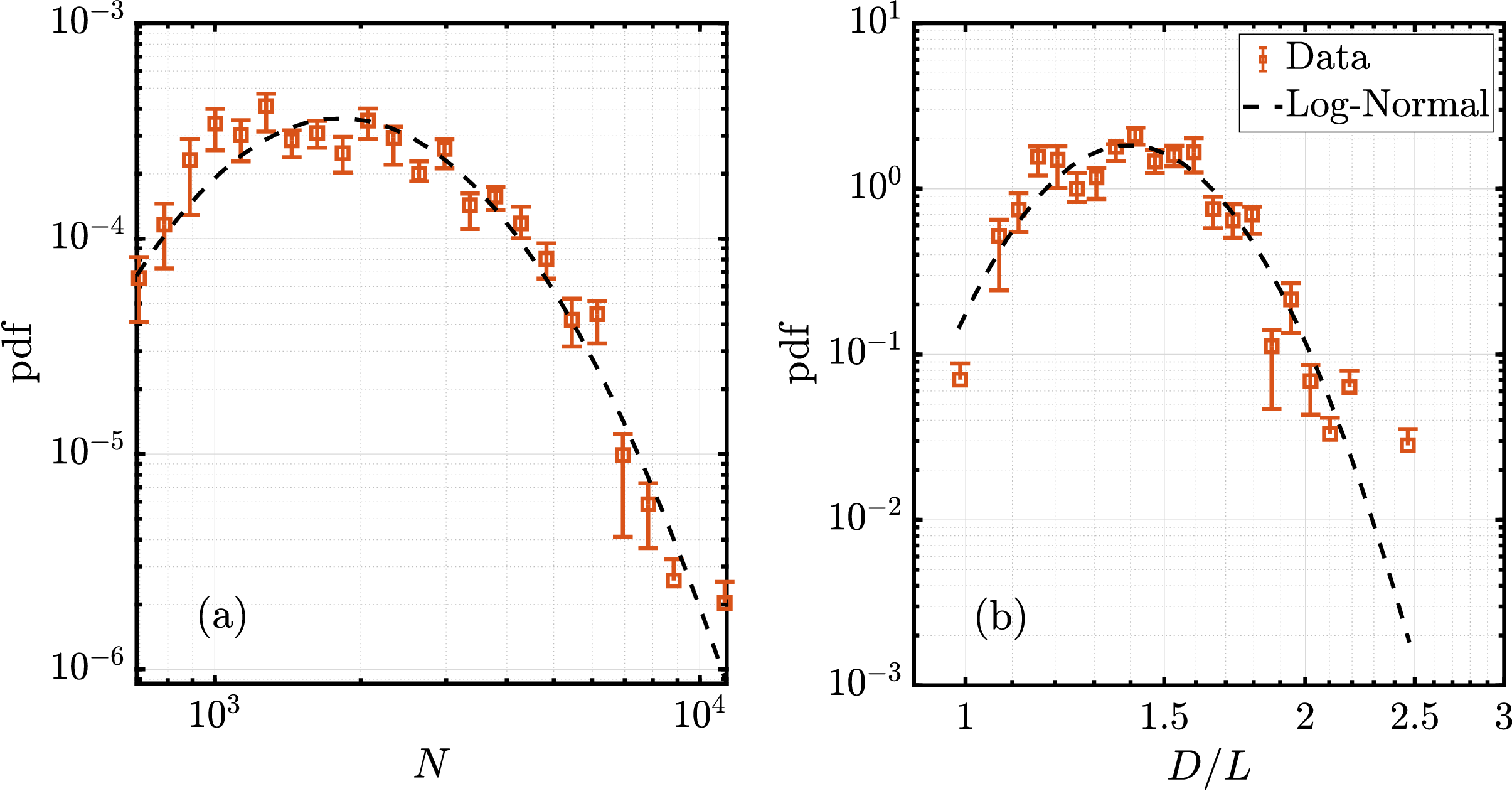}
    \caption{Distribution of the number of fibres per aggregate (a) and of the diameter of the aggregate (b).
    The dashed lines, correspond to a log-normal distribution having the same mean and standard deviation than the measured ones. 
    Error bars are estimated here by considering 20 random sets of 80\% of the data and by calculating the histogram for each set, which gives a minimum and maximum deviation for the points.
    These data are from a set of aggregates formed at $f = \SI{9}{\hertz}$ with a volumic fraction $\varphi = \SI{1.32e-2}{}$.
    The experiment lasted for $T = \SI{5}{h}$.}
    \label{fig:distribution}
\end{figure}

The distributions of the number of fibres per aggregate and of its diameter are shown on Figure~\ref{fig:distribution} for $f = \SI{9}{\hertz}$ and $T= \SI{5}{h}$ (other frequency and duration give similar results). 
Both distributions are compatible with a log-normal distribution.
This is consistent with the expectation that if $Y$ follows a log-normal distribution and $Y=X^\alpha$, then $X$ also follows a log-normal distribution.
This behaviour has also been observed in the mass and size distributions of aegagropilae, natural fibre aggregates \cite{verhille_structure_2017}.
The log-normal distribution is commonly observed in aggregation and fragmentation processes where the growth or fragmentation rate varies with time but is size-independent \cite{koch_logarithm_1966}. 
These processes can be described by:
\begin{equation}
    \dv{N}{t} = k(t) N 
\end{equation}

where $k(t)$ is the growth or the fragmentation rate, independent of $N$ but fluctuating randomly over time.

\begin{figure}[htbp]
    \includegraphics[width=1\linewidth]{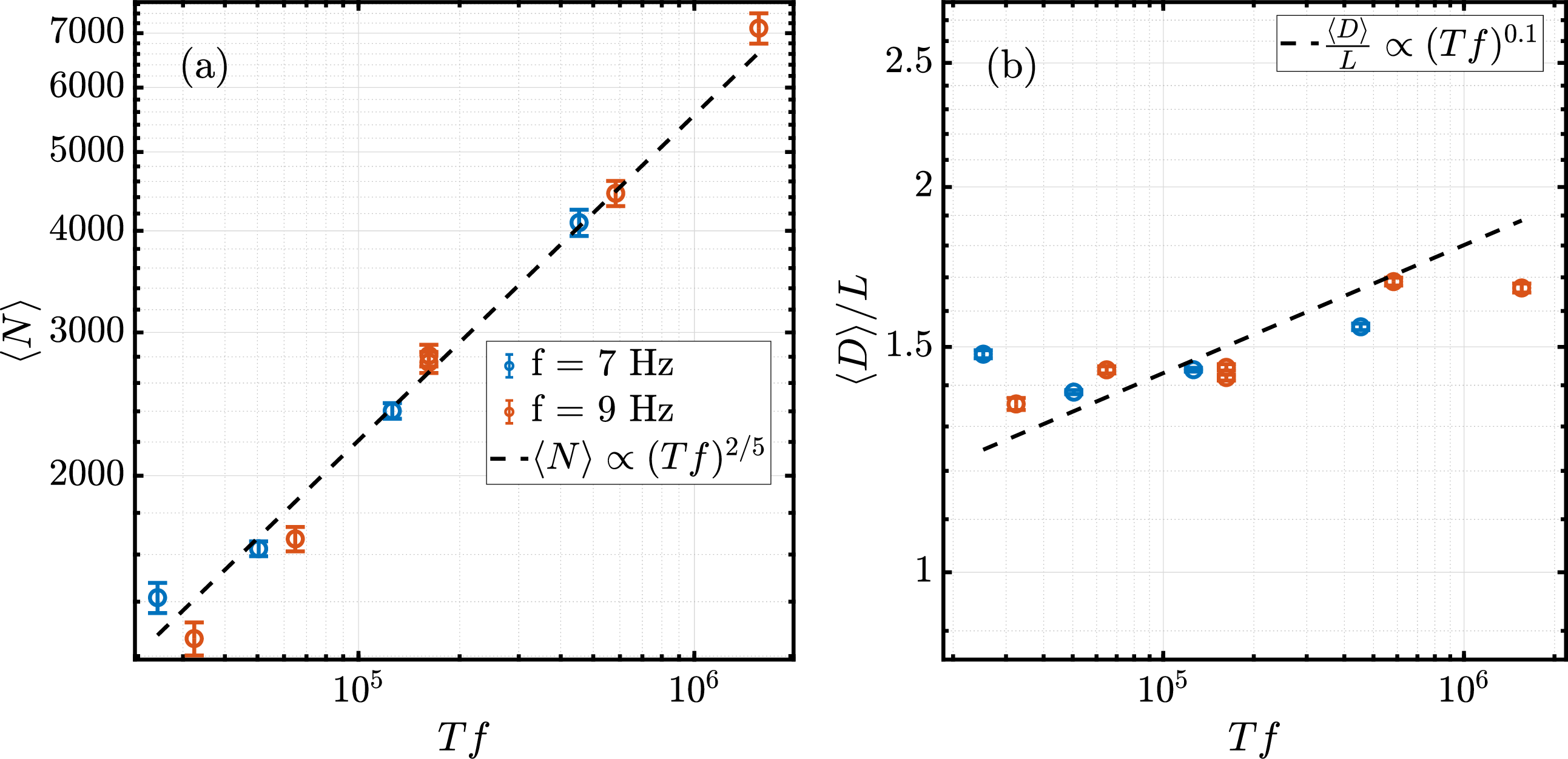}
    \caption{Evolution of the mean number of fibres per aggregate (a) and the mean aggregate diameter (b) as a function of $Tf$ which corresponds to the number of disc rotation periods.
    The error bars are calculated from the standard deviation of the distributions. 
    They represent the natural dispersion of the data in each experiment but do not account for the variability corresponding to the reproducibility of the experiment.
    One experiment was repeated, corresponding to the data points at $Tf =\SI{1.6e5}{}$ that stay very close one to the other.}
    \label{fig:NRmean}
\end{figure}

The evolution of the mean number of fibres per aggregate $\langle N\rangle$ as a function of $T f$ is presented in Figure~\ref{fig:NRmean}(a).
The quantity $T f$ represents the total number of disc rotations or equivalently the number of integral times, and can be interpreted as the average age of the aggregates.
$\langle N\rangle$ is observed to follow the scaling law $\langle N\rangle\propto\qty(T f)^{2/5}$.
However, no explanation have yet been found for this scaling which is observed at large time compared to the typical timescales of the flow.
By extrapolating the trend to $T f=1$, corresponding to one turbulent integral time, one can find $\langle N\rangle|_{T f = 1} = 18$.
This suggests that, within one disc rotation period, a sufficient number of fibre-fibre collisions can occur to form a coherent structure.
Indeed, it has been shown that only five contacts are enough to interlock fibres~\cite{switzer_flocculation_2004}.

In Figure~\ref{fig:NRmean}~(b), the mean aggregate diameter is plotted as a function $Tf$.
\refun{Although the data exhibit significant dispersion, the aggregate radius shows a slight increase over time. 
In Figure~\ref{fig:NRmean}~(b), a power law in $(Tf)^{0.1}$ is also plotted, which is the expected scaling based on the previously established experimental relationships: $N \propto D^4$ and $\langle N \rangle \propto (Tf)^{2/5}$.
The discrepancy between the power law and the experimental data can likely be attributed to the dispersion of the values of $D$: since $D$ is broadly distributed, it is generally not valid to assume that $\langle D^n \rangle = \langle D \rangle^n$.}

Finally, no clear correlation was found between the number of aggregates formed and the experimental input parameters $T$, $f$ and $\varphi$.

%%%%%%%%%%%%%%%%%%%%%%%%%%%%%%%%%%%%%%%%%%%%%%%%%%%%
%%%%%%%%%%%%%%%%%%%%%%%%%%%%%%%%%%%%%%%%%%%%%%%%%%%%%
%%%%%%%%%%%%%%%%%%%%%%%%%%%%%%%%%%%%%%%%%%%%%%%%%%%%%

\section{\label{secFrag}Fragmentation}

%%%%%%%%%%%%%%%%%%%%%%%%%%%%%%%%%%%%%%%%%%%%%%%%%%%%
\subsection{\label{secFragtime}Fragmentation time}
\begin{figure}[htbp]
\centering
 \includegraphics[width=0.5\linewidth]{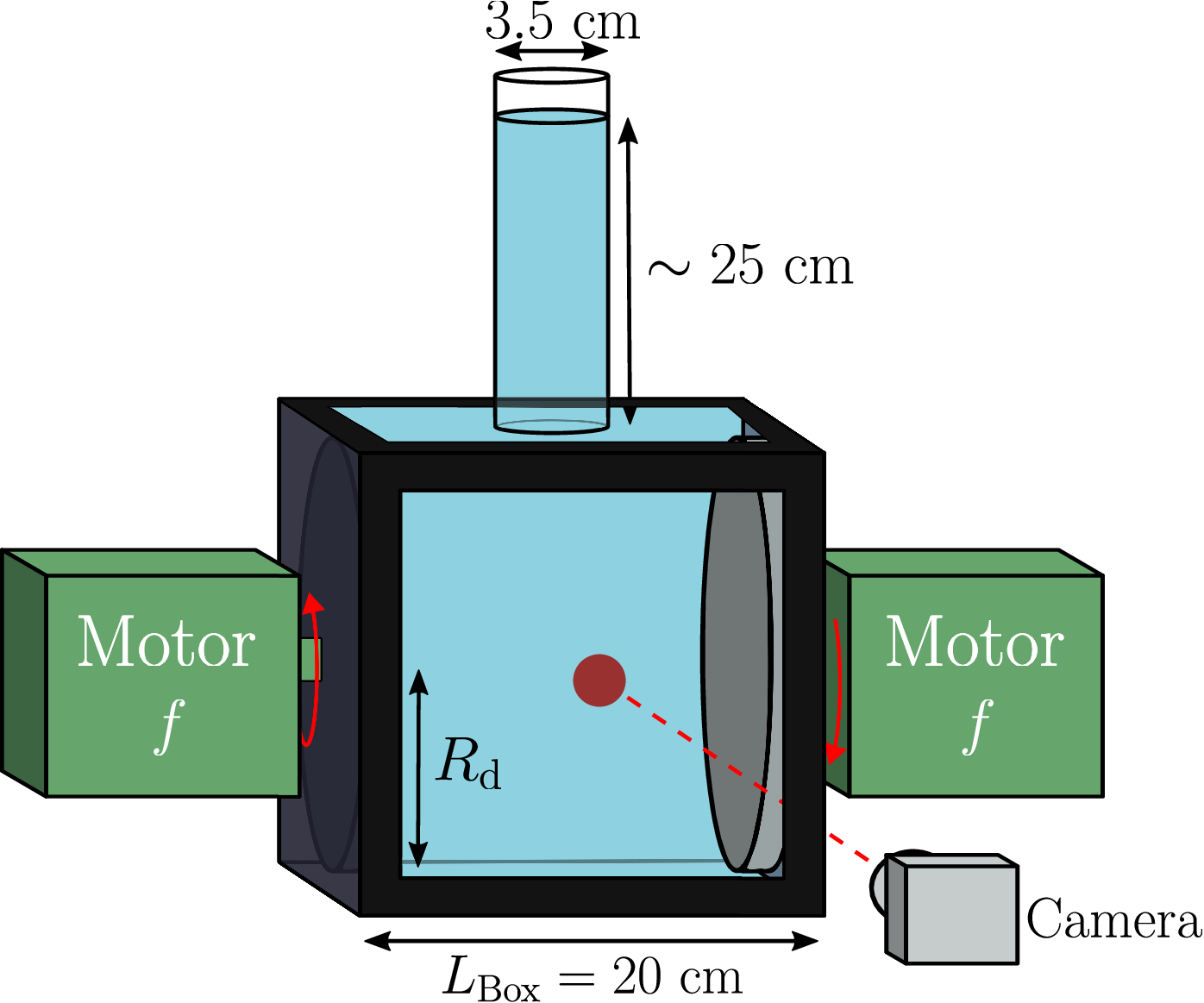}
   \caption{Sketch of the modified setup used to investigate fragmentation. The radius of the discs $R_d =\SI{8}{\centi\meter}$ and the size of the tank $L_\text{Box} = \SI{20}{\centi\meter}$ is unchanged. Only a chimney has been added.}
    \label{fig:Setupmod}
\end{figure}
\begin{figure}[htbp]
    \includegraphics[width=1\linewidth]{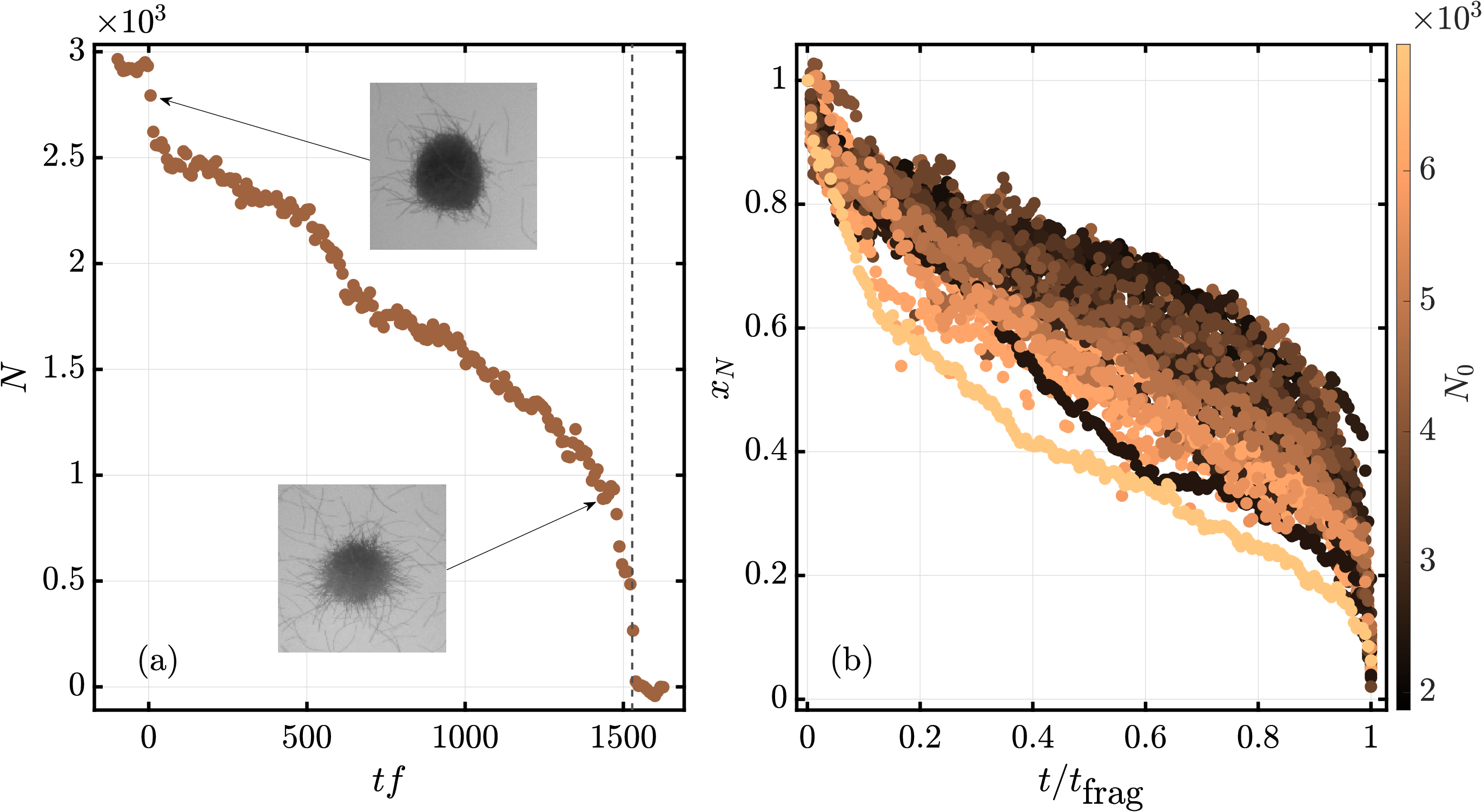}
    \caption{(a) Number of fibres in the aggregate as a function of time. The images correspond to snapshots of the aggregate during the fragmentation process. The size of the picture is approximatively \SI{3.7}{\centi\meter}. The dashed line indicates the time at which the aggregate is fully destroyed: $t_\text{frag}f$. (b) Evolution of $x_N = N/N_0$, the fraction of fibres remaining in the aggregate as function of the time rescaled by the fragmentation time. The color map represents $N_0$, the initial number of fibres in the aggregate.}
    \label{fig:exfrag}
\end{figure}

As the growth rate of the aggregates is expected to result from a balance between the aggregation of new fibres and the erosion of already aggregated fibres due to hydrodynamic stress, the setup has been modified to investigate only the fragmentation process.
An access point was added to the tank, allowing to introduce a single aggregate into the flow while the experiment was running, as shown on Figure~\ref{fig:Setupmod}(a).
The added chimney is flush on the top of the tank and it is assumed that the flow properties are mainly unaffected by this modification.
In this configuration, the tank is filled with water without fibres.
A single aggregate is first rehydrated and introduced into the von Kármán turbulent flow.
A backlight is used so that the aggregates and the eroded fibres appear black against the background. 
During the fragmentation process, the number of fibres remaining in the aggregate is tracked by analysing the background images. 
When the fibres are eroded from the aggregate, they remain suspended in the tank, gradually darkening the background.
The luminosity is assumed to be linearly dependent on $N$, the number of fibres remaining in the aggregate. 
This is a reasonable assumptions in the range of fibres corresponding to one aggregate.
Therefore, recording the luminosity of the images in time allows the measurement of the evolution of $N$ the number of fibres in an aggregate.
The evolution $N$ as a function of $tf$ is shown Figure~\ref{fig:exfrag} where $t$ is the current time and $f$ the frequency of the disc rotation.
A first observation is that, over time, fibres are slowly eroded from the aggregate until it is fully destroyed, releasing all the remaining fibres of a much smaller time scale.
This fragmentation process does not correspond to a single fragmentation event that separate all the fibres instantaneously. 
Instead, fibres are released gradually over a long timescale compared to the characteristic timescale of the flow $1/f$.
The fragmentation time $t_\text{frag}$ is defined as the time needed to fully destroy the aggregate as shown with the dashed line on Figure~\ref{fig:exfrag}~(a).
The same figure also includes two zoomed snapshots of the aggregate during the fragmentation process.
The aggregate keeps its spherical shape even when more than $2/3$ of the fibres have been eroded.

Figure~\ref{fig:exfrag}~(b) presents the evolution of $x_N = N/N_0$, the fraction of fibres remaining in the aggregates as function of the time $t$ rescaled by the fragmentation time $t_\text{frag}$ for different aggregates. 
All fragmentation processes are observed to follow a similar trend: 
A slow erosion phase lasting approximately 90\% of the total fragmentation time, followed by a final breakup where the last \qtyrange{20}{40}{\%} of the fibres are released.

\begin{figure}[htbp]
\centering
    \includegraphics[width=0.8\linewidth]{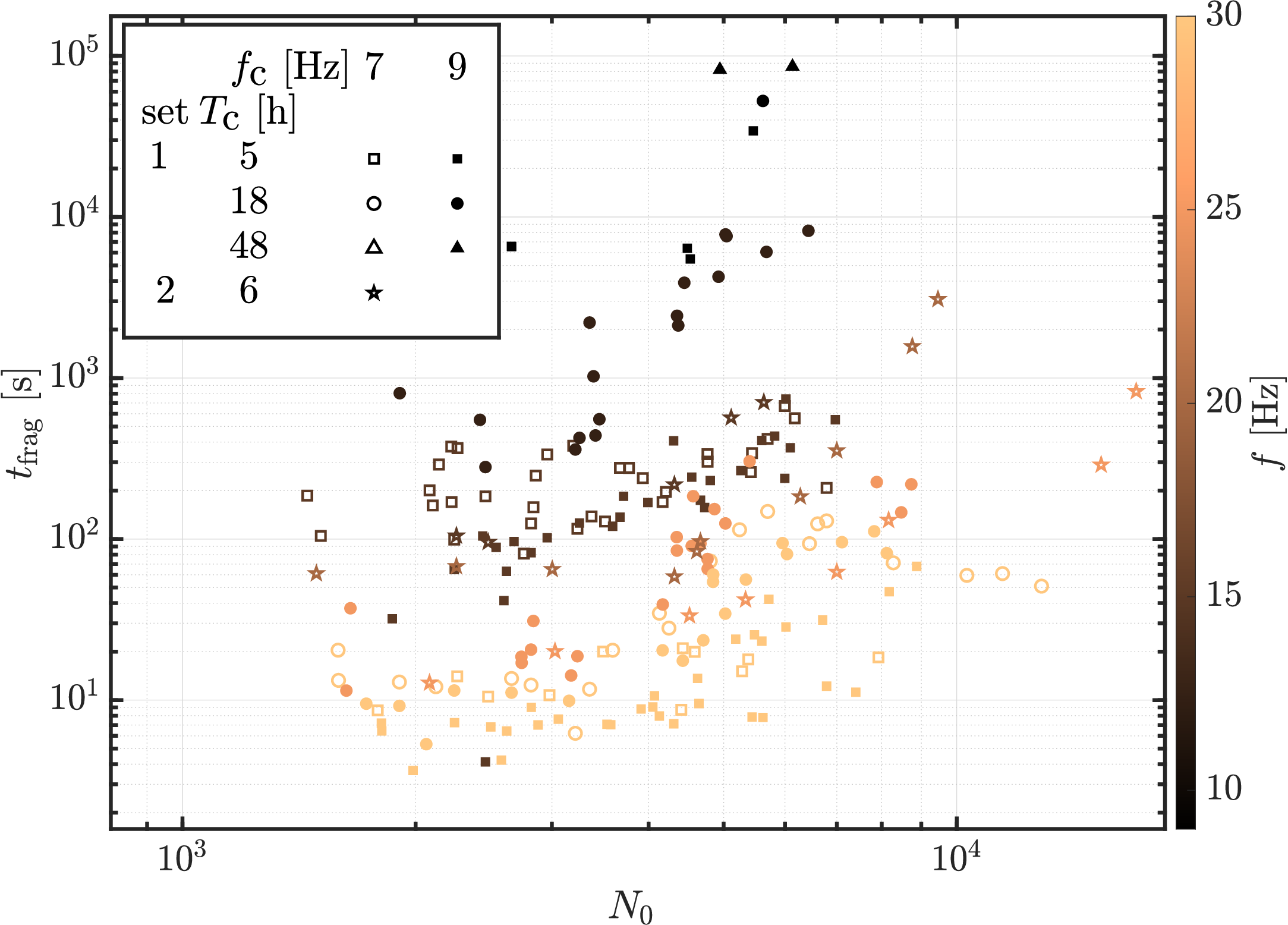}
    \caption{\reftrois{Fragmentation time $t_\text{frag}$ as a function of the number $N_0$ of fibers in the aggregate. Each point corresponds to a fragmentation experiment with a different aggregate. The color code represents the frequency used for fragmentation, and the symbols indicate the formation parameters. In this figure, all aggregates from set 1 were formed with an initial volume fraction of free fibers of $\varphi = \SI{1.32e-2}{}$, and $\varphi = \SI{1.56e-2}{}$ for set 2.}}
    \label{fig:tfrag}
\end{figure}

The evolution of fragmentation times of the aggregates, $t_\text{frag}$, is shown in Figure~\ref{fig:tfrag} as a function of $N_0$, the number of fibres in the aggregate.
First, note the data points span more than 4 orders of magnitude which corresponds to the influence of the experimental parameters.
Moreover a natural dispersion due to the natural diversity of the aggregates is observed. 
Note that the conditions of the aggregates formation (in terms of $f_\text{c}$ and $T_\text{c}$) does not significantly influence $t_\text{frag}$, within the tested parameter range.
$t_\text{frag}$ is observed to increase with $N_0$.
This is expected, as larger aggregates contain more fibres to disentangle.
Additionally, since larger aggregates have been found to be denser (see Section~\ref{Characterisation}), they are expected to be more resistant to fragmentation.
Furthermore, $t_\text{frag}$ decreases as the excitation frequency increases.
This is consistent with the fact that hydrodynamic stresses become stronger as the disc frequency increases, leading to faster aggregate breakups. 
However with a simple scaling model, one could have expected the fragmentation time to be inversely proportional to the power of turbulence at the scale of the aggregate: $t_\text{frag} \propto 1/(\rho R^3 \epsilon) \propto 1/f^3$.
Unfortunately, this scaling approach fails to collapse the experimental data on a single curve. 
As it will be presented now, a deeper understanding of the fragmentation process is required to accurately model it.

\begin{figure}[htbp]
\centering
    \includegraphics[width=0.8\linewidth]{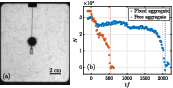}
   \caption{(a) Snapshot of an aggregate held by a needle during a fragmentation experiment.(b) Temporal evolution of the number of fibres within two similar aggregates during fragmentation. The blue curve represents an aggregate held by a needle, while the red curve corresponds to a free-moving aggregate. The measurements for each case are taken separately. The dotted lines indicate the fragmentation times. Despite having a similar initial number of fibres, the fixed aggregate fragmentation lasts four times longer than the free one.}
    \label{fig:aig}
\end{figure}

Fragmentation is complex as it depends on both the structure of the network of fibres within the aggregate and on the forces acting on it.
In the set up, during a fragmentation experiment, an aggregate is free to move throughout the tank and it collides intermittently with the discs.
\refun{This effect cannot be avoided in a finite size experiment for a free aggregate with a lifetime much longer than the integral time of turbulence.}
To isolate the effect of the hydrodynamical forces, aggregates are held at the centre of the flow using a needle, as shown on Figure~\ref{fig:aig}~(a).
In this configuration the aggregate is not advected by the flow anymore, and the collisions with the discs are prevented. 
Under these conditions, the fragmentation time increases by a factor 3 to 10 compared to free aggregates as shown on Figure~\ref{fig:aig}~(b).
This suggests that even if hydrodynamical forces cannot be totally ignored, collisions with the discs facilitate fibres rearrangement and subsequent erosion of the aggregate.
\refun{Two phenomena seem to be at play. First, rearrangement of the network caused by collisions in the case of free aggregates, and by turbulent pressure fluctuations for fixed aggregates. Second, rapid erosion of the loosely entangled fibres, which leads to their dispersion. The limiting timescale corresponds to the rearrangement of the network.}

\begin{figure}[htbp]
\centering
    \includegraphics[width=0.5\linewidth]{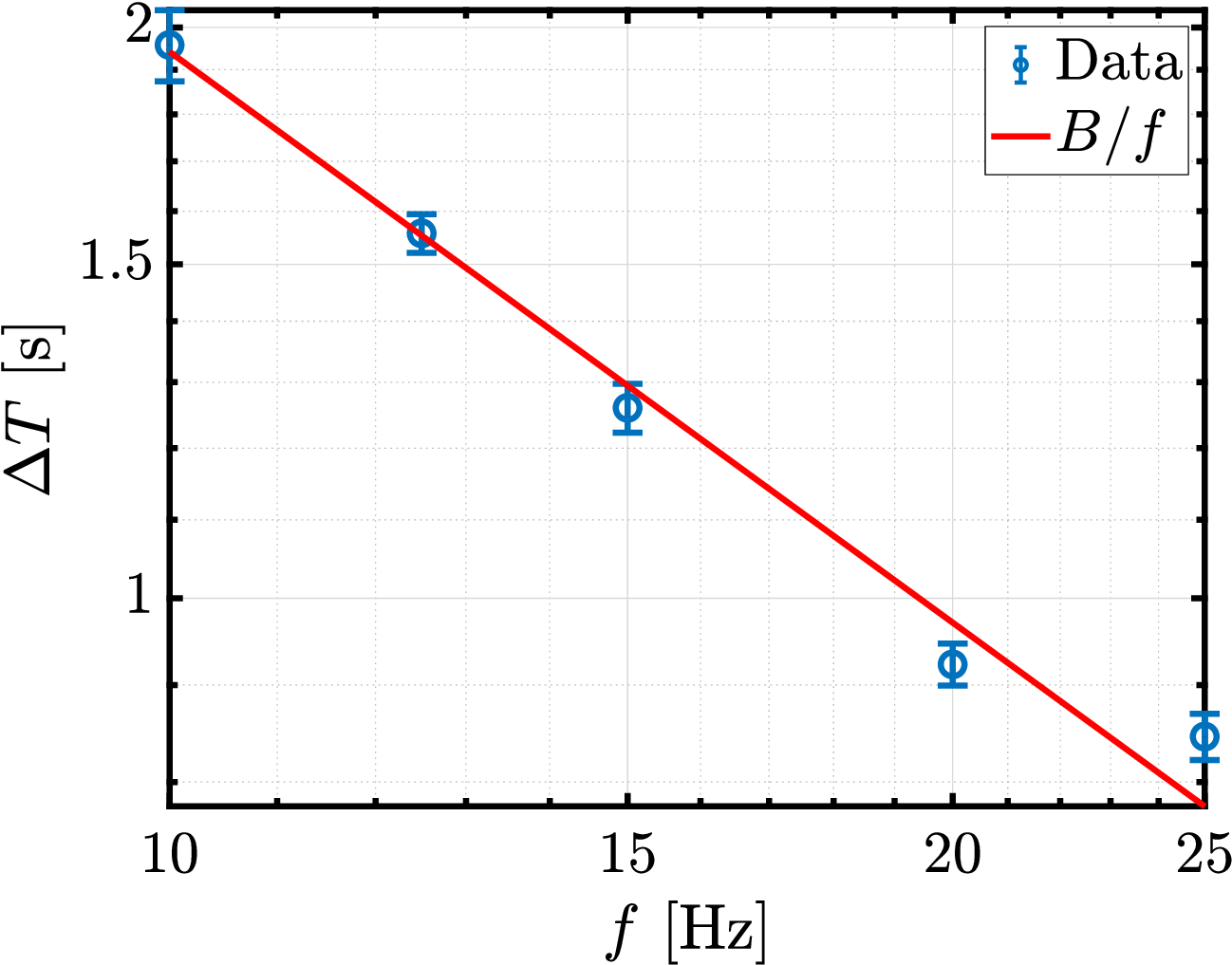}
    \caption{Mean time between collisions of aggregates with the discs as a function of the imposed motor frequency. The red line corresponds to the best fit of the form $\Delta T = B/f$, with $B = 19.4$. The collisions are spotted by eye. Each point corresponds to an average over 4 fragmentation processes with between \qtyrange{50}{500}{} collision events in each.}
    \label{fig:Contactoeil}
\end{figure}

To better understand fragmentation dynamics that occurs in the setup, the role of collisions between aggregates and the discs is investigated.
First, for free aggregates, the collisions between the aggregates and the discs are visually identified.
The mean time between this collisions are presented on Figure~\ref{fig:Contactoeil}.
The observed scaling behaviour is of the form $\Delta T \propto 1/f$.
This scaling is discussed in more detail in the Appendix~\ref{App}

%%%%%%%%%%%%%%%%%%%%%%%%%%%%%%%%%%%%%%%%%%%%%%%%%%%
\subsection{\label{toymodel}A Model to predict the fragmentation time}

We have found experimentally that the collisions with the discs play a central role in the fragmentation process.
Furthermore, one can expect that the collisions induce structural reorganisation of the fibres that allows the fibres to be extracted more easily from the aggregate.
Therefore we propose a model where the timescale of fragmentation, which is much larger than the integral time, is determined by the number of collision with the discs, $N_\text{col}$. 
This number of collisions is assumed to be given by the ratio of the elastic energy $\mathcal{E}_\text{el}$ stored by the fibres within the network of an aggregate and the energy $W$ gain by the aggregate during a single collision with a disc:

\begin{equation}
     N_\text{col} = \frac{\mathcal{E}_\text{el}}{W}
    \label{mastereq}
\end{equation}

Furthermore $W$ can be estimated as $W = F \cdot \Delta x$, where $F = \rho v^2 C_x \pi D^2/8$
is the hydrodynamic force on the aggregate near the disc with $C_x$ the drag coefficient and $\Delta x$, the aggregate deformation during a collision.
$v$ is taken as $v\sim R_\text{d}f$ with $R_\text{d}$ being the radius of a disc. 
Moreover, the deformation $\Delta x$ of the aggregate of diameter $D$ submitted to a force $f$ can be estimated using an Hertzian contact~\cite{lifshitz_copyright_1986}:
\begin{equation}
    \Delta x = \frac{1}{2}\left(\frac{3F^2}{E_\text{eq}^2 D}\right)^{1/3}
\end{equation}

Here, $E_\text{eq} = A E \phi^3$ is the aggregates equivalent Young's modulus~\cite{verhille_structure_2017,van_wyk_20note_1946,gey_structural_2025}, with $E$ the fibre Young's modulus, $\phi$ the aggregate volume fraction and $A$ a numerical constant of the order of $A \sim \num{1e-1}$ .

This allows to rewrite $W$ as :
\begin{equation}
    W \propto \frac{3^{2/3}}{64}\qty(\frac{\pi^5 C_x^5 \rho^5 v^{10}}{A^2 E^2})^{1/3}\frac{D^3}{\phi^2}
    \label{W}
\end{equation}

On the other hand, the elastic energy required to extract $N$ fibres from the network scales as $\mathcal{E}_\text{el} = L E I \kappa^2 N$, where $L$ is the length of a fibre, $I = \pi d^4/64$ is the moment of inertia of the fibre of diameter $d$ and $\kappa$ is the curvature of extracted fibres.

Moreover, the number of collisions before the aggregate is fully dispersed can be written as $N_\text{col} = t_\text{frag} / \Delta T$ with $\Delta T = B / f$ and $B\sim20$ (see Figure~\ref{fig:Contactoeil}).

Equation~\ref{mastereq} leads to:
\begin{equation}
    t_\text{frag} \propto \frac{C}{f^{13/3}}\qty(\frac{\kappa^2N\phi^2}{D^3}) \qq{with} C = \frac{BA^{2/3}LE^{5/3}d^4}{(3\pi)^{2/3}C_x^{5/3}\rho^{5/3}R_\text{d}^{10/3}}
    \label{Semiresol}
\end{equation}

\begin{figure}[htbp]
\centering
    \includegraphics[width=0.5\linewidth]{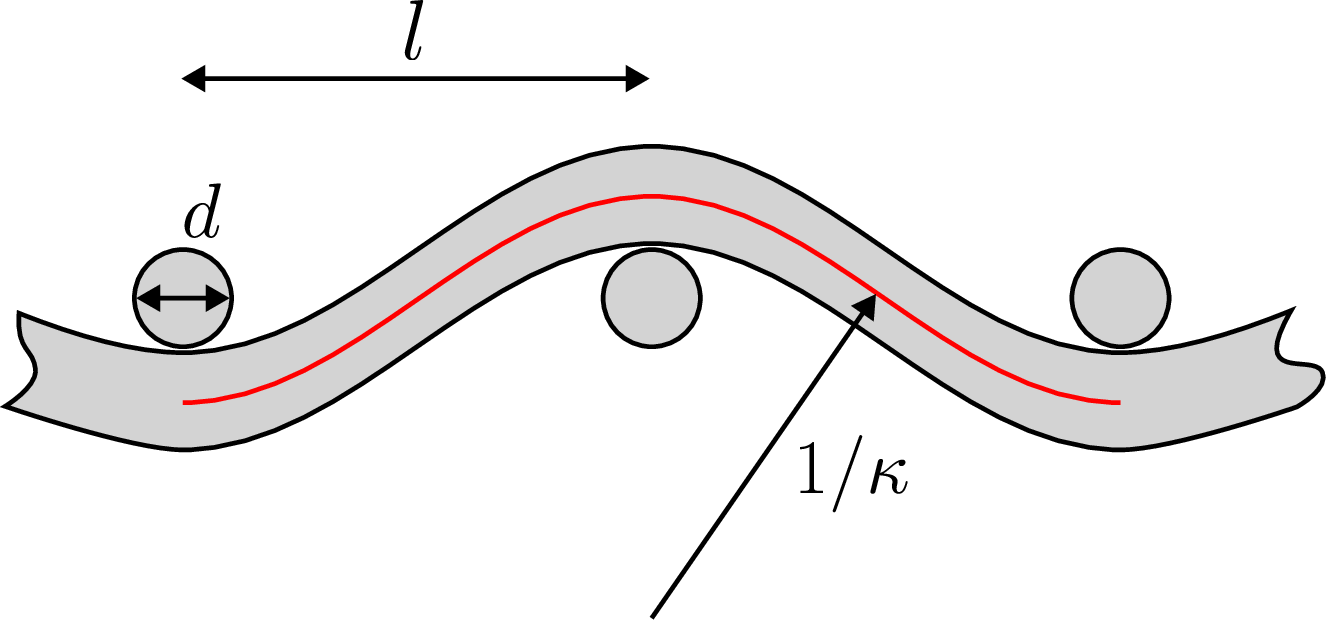}
   \caption{Sketch of the curvature of a fibre. $d$ is the diameter of a fibre and $l=L/N_{C}$ the typical distance between contacts points.}
    \label{fig:fibre}
\end{figure}

\reftrois{To go further, the scaling law observed for the aggregates compaction is used: $N \propto D^4$ (see Section~\ref{Characterisation}).
The nondimensionalisation of $D$ is chosen based on a fractal dimension analysis that accounts for the fiber volume~\cite{de_la_rosa_zambrano_fragmentation_2018}: $l = (Ld^2)^{1/3}$, so that $N = \alpha (D/l)^4$ with $\alpha \sim \num{1.0e-3}$.
The choice of this lengthscale has been made to recover the right evolution of the fragmentation time with the fiber aspect ratio $\Lambda$, observed in Figure~\ref{fig:tfrag}. 
Nevertheless, such a change does not affect the scaling with respect to $N$ and $f$.}
To estimate $\kappa$, the curvature required for fibres to escape from the network, a typical situation as illustrated in Figure~\ref{fig:fibre} is considered.
The order of magnitude for $\kappa$ can be estimated as $\kappa \sim {2d}/\qty(l^2 + d^2) \sim {2\phi^2}/{d}$, where $l$ the distance between contact points is derived from the scaling for random rod packing~\cite{philipse_random_1996} and leads to $l = {L}/{N_{C}} = {d}/{\phi}$, with $N_{C}$ being the mean number of contact points per fibre.
Note that this estimation is consistent with the fact that $E_\text{eq}\sim E\phi^3$.

Combining these results with Equation~\ref{Semiresol}, leads to:

\reftrois{
\begin{equation}
    \frac{t_\text{frag}}{t_\text{th}} \propto N^{7/4}\qq{and} t_\text{th} = A'\qty(\frac{E}{\rho R_\text{d}^{2}})^{5/3}\frac{1}{f^{13/3}}
    \label{eq.final}
\end{equation}}

\reftrois{where $A' = 2^7B\alpha^{21/4}\qty[A^2/\qty(3^5\pi^2C_x^5)]^{1/3}$ is a numerical constant, found to be $A' \sim \num{1.2e-13}$ (with $C_x \sim 0.5$, which corresponds to a typical order of magnitude for the drag coefficient in the aggregates turbulent regime).}

\begin{figure}[htbp]
\centering
    \includegraphics[width=0.8\linewidth]{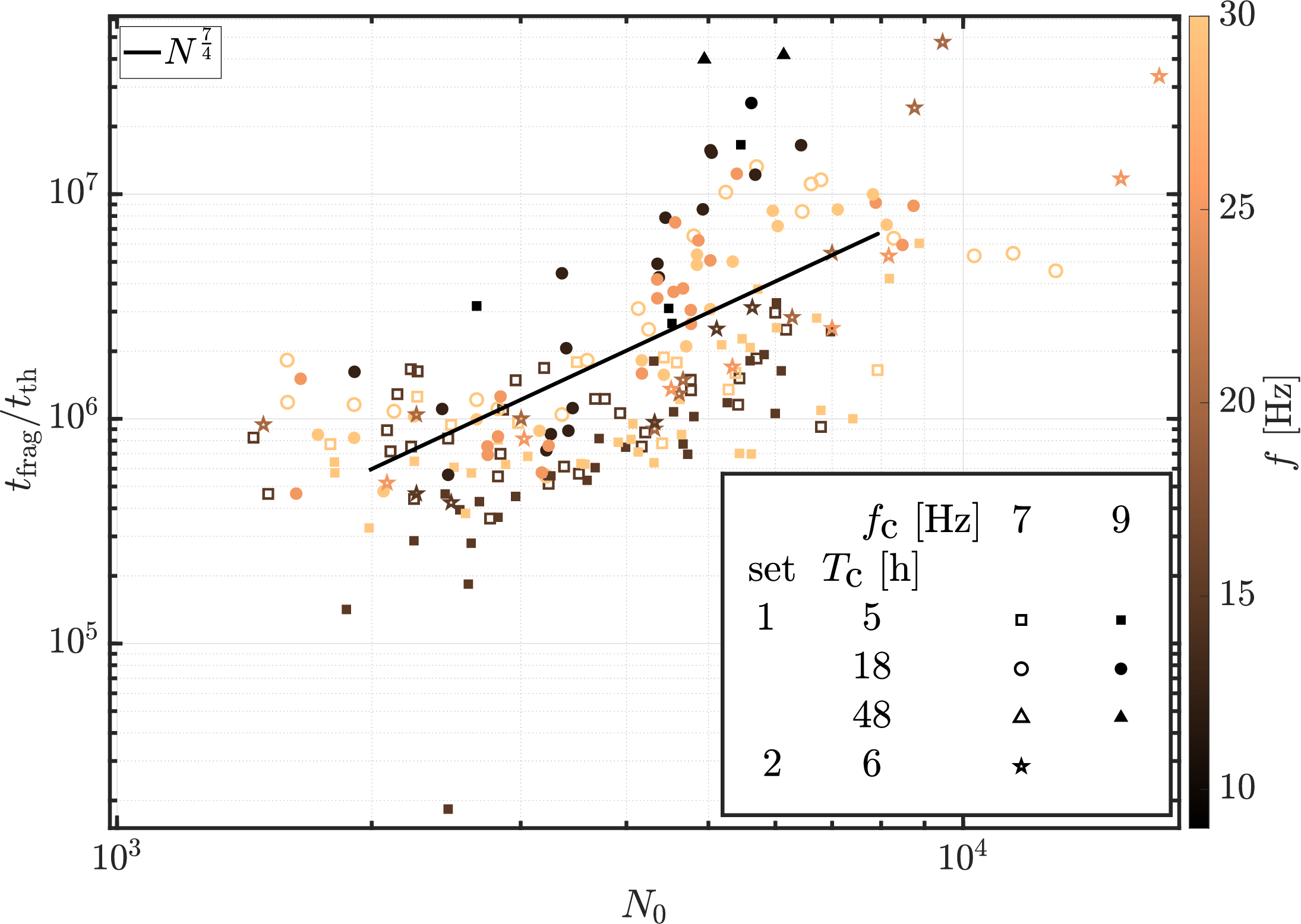}
    \caption{\reftrois{Fragmentation time $t_\text{frag}$ as a function of the number of fibers in the aggregate, normalized by $t_\text{th}$ given by Equation~\ref{eq.final}.
Each point corresponds to an experiment with a different aggregate. The color code represents the excitation frequency used for fragmentation, and the symbols indicate the formation parameters.
In this figure, all aggregates from set 1 were formed with an initial volume fraction of free fibers $\varphi = \SI{1.32e-2}{}$, and $\varphi = \SI{1.56e-2}{}$ for set 2.
The black line corresponds to the result of the fragmentation model: $t_\text{frag}/t_\text{th} = N^{7/4}$.}}
    \label{fig:rescale}
\end{figure}

Applying this scaling to the measurements of the fragmentation time presented in Figure~\ref{fig:tfrag}, permits to decrease the dispersion of the data by a factor 100 (see Figure~\ref{fig:rescale}). 
Although the collapse is not perfect, as the natural diversity of the aggregate population remains, the experimental data are consistent with the observed $N^{7/4}$ power law.

\refun{The model presented here highlights the role of collisions between the discs and the aggregates in the fragmentation process.
However, while collisions increase the rate at which the network rearranges, they are not the only mechanism by which such rearrangement can occur. 
As shown in Section~\ref{secFragtime}, the aggregates can break even in the absence of collisions with the discs. 
This suggests that pressure fluctuations could also reorganize the fibre network with a lower rate.
Nevertheless, this turbulent rate has yet to be determined, as free aggregates in wall-free turbulence remain challenging to investigate experimentally.}

%%%%%%%%%%%%%%%%%%%%%%%%%%%%%%%%%%%%%%%%%%%%%%%%%%%%
%%%%%%%%%%%%%%%%%%%%%%%%%%%%%%%%%%%%%%%%%%%%%%%%%%%%%%%
%%%%%%%%%%%%%%%%%%%%%%%%%%%%%%%%%%%%%%%%%%%%%%%%%%%%%

\section{Conclusion}

In this paper, an experimental setup to create and study fibre aggregation in a turbulent flow have been presented. 
The aggregates are formed in a range were the volume fraction is high enough to be in a concentrated regime~\cite{kerekes_characterization_1992}.
Contrary to previous experiments~\cite{kerekes_characterization_1992,soszynski_elastic_1988}, the turbulence intensity plays a role in the appearance of aggregates.

The distributions of sizes and the number of fibres within an aggregate are compatible with log-normal distributions, which is commonly observed in aggregation or fragmentation processes when the rate is independent of the aggregated object's size~\cite{koch_logarithm_1966}.

The joint distribution of sizes and number of fibres also suggests that the formation process of the aggregates is associated with compaction, leading to a scaling law of the form $N\propto (D/L)^4$.
The mean number of fibres per aggregate is described by $\langle N\rangle \propto \qty(T\times f)^{2/5}$, showing the growth of the aggregate with their age.

Then, focusing solely on fragmentation, the influence of collision with the discs are examined and a model to explain the fragmentation time as a function of the imposed frequency of the von Kármán rotating discs and the number of fibres in the aggregate is proposed. 
Although this model cannot predict the temporal evolution of the aggregates erosions, it gives a reasonable estimation of the fragmentation time.

Additionally, by considering only the time dependence, the scaling laws for both aggregation and fragmentation are consistent with a competition between these two processes taking place in the formation of the aggregate.
Specifically, the fragmentation curve $T\propto N^{7/4}$ lies below the aggregation curve $T\propto N^{5/2}$ for small $N$, and the reverse is true for large $N$.
This implies that it takes longer to break a aggregate than to form it at small $N$, so the aggregates should continue to grow until the two curves intersect.

\refdeux{The main findings presented in this article can be summarized as follows:
\begin{itemize}
\item Dense fibre aggregates are formed in a turbulent flow. They are observed when the turbulence intensity is high enough to mix the fibres and enhance collisions between them, but low enough so that the hydrodynamic forces are not too strong, preventing the dispersion of aggregate.\\
\item The aggregates are found to undergo compaction during their formation process, leading to a scaling of the form $N \propto D^4$.\\
\item The number of fibres in the aggregates is compatible with a log-normal distribution, characteristic of growth or fragmentation processes with a random rate independent of $N$.\\
\item The fragmentation process of the aggregates corresponds to a slow erosion of the aggregates for 90\% of the time, followed by a final rapid breakup.\\
\item By taking into account the collisions between the aggregates and the discs, a model for the fragmentation time is proposed. The main physical idea is that collisions allow the reorganisation of the fibres, which can then be eroded by turbulence.
\end{itemize}
}

\appendix

\section{\label{App} Role of disc collisions in fragmentation}

To better understand fragmentation dynamics that occurs in the setup, the role of collisions between an object and the discs is investigated.

\begin{figure}[htbp]
\includegraphics[width=1\linewidth]{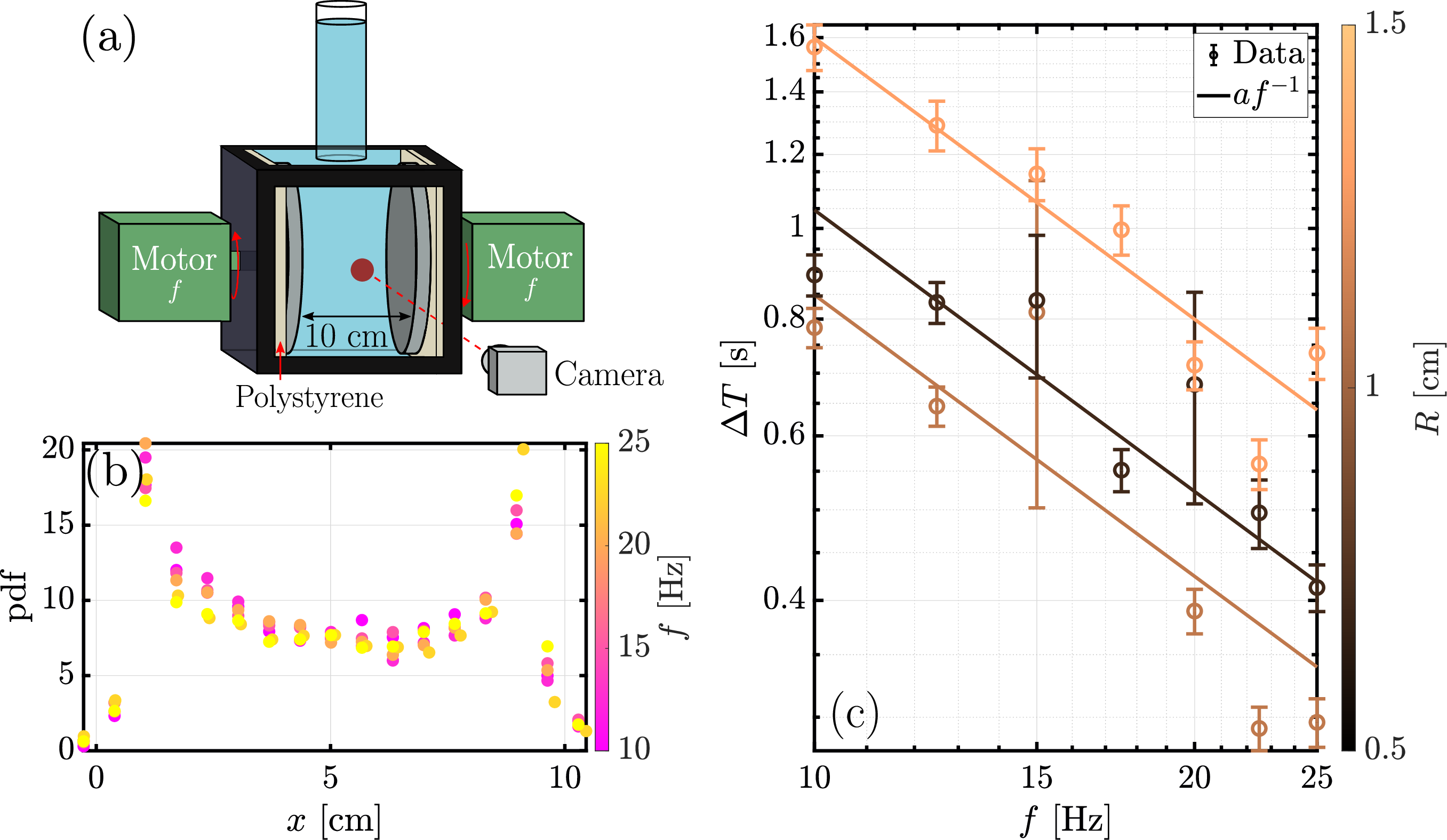}
    \caption{(a) Sketch of the modified setup to measure the collision between the sphere and the discs. The discs are brought closer to one another so that the distance between the discs is \SI{10}{\centi\meter} and that they are visible on the camera. (b) Distribution of the positions of the sphere along the inter-disc direction for a sphere of radius \SI{1}{\centi\meter}. (c) Time between collisions with the discs as a function of the imposed motor frequency for neutrally buoyant spheres. The colour scale corresponds to the radius of the spheres. $a=\SI{11\pm4}{}$.}
    \label{fig:Contact}
\end{figure}
For free aggregates, a scaling behaviour of the form $\Delta T \propto 1/f$ is found (Figure~\ref{fig:Contactoeil}).
However, during the fragmentation process, the aggregates are eroded, and eventually broken.
To go further and characterize the collision rate independently of the fragmentation and elasticity of the aggregate, neutrally buoyant plastic spheres with radii ranging from \SIrange{0.5}{2}{\centi\meter} were used.
Moreover, in the original set up, the discs are not visible by the camera.
To automate the measurement, the setup is modified to reduce the distance between the discs, allowing direct visualization, as shown of Figure~\ref{fig:Contact}(a).
Polystyrene panels are added behind the discs to prevent the spheres to go behind the discs.
This leads to a smaller von Kármán turbulent flow with a \SI{10}{\centi\meter} gap between the discs and \SI{20}{\centi\meter} in the other directions.
Using spheres allowed, after a calibration, to obtain the 3D trajectories of the spheres using a single camera, as the distance from the camera is encoded in the apparent size of the spheres.
This set up is inspired by previous work by Machicoane et al. who observed a preferential trapping of large spheres close to the discs within a turbulent von Kármán flow~\cite{machicoane_dynamique_2013}.

Figure~\ref{fig:Contact}(b) shows the spatial distribution of the spheres positions. 
Spheres spend more time near the discs than at the tank centre, which is consistent with prior observations~\cite{machicoane_dynamique_2013}.
Large particles, relative to the Kolmogorov scale, in a von Kármán turbulent flow, are preferentially trapped near the discs.
This positional bias underlines the relevance of disc collisions in the fragmentation process.

By identifying the collisions events, $\Delta T$, the time between collisions is measured (Figure~\ref{fig:Contact}(c)).
The scaling found is $\Delta T \propto 1/f$, with a scaling factor independent of the size of the sphere in the range we tested (always much larger than the Kolmogorov length), confirming the result of Figure~\ref{fig:Contactoeil} previously obtained for fibre aggregates.

%%%%%%%%%% Insert bibliography here %%%%%%%%%%%%%%

%\bibliographystyle{RS} %%%% .BST file

\bibliography{bib.bib}

\end{document}